\documentclass[acmlarge]{acmart}

\AtBeginDocument{%
  \providecommand\BibTeX{{%
    \normalfont B\kern-0.5em{\scshape i\kern-0.25em b}\kern-0.8em\TeX}}}

\setcopyright{acmcopyright}
\copyrightyear{2023}
\acmYear{2023}
\acmDOI{nn.nn}

\acmConference[CHI '23]{Workshop on Designing Technology and Policy Simultaneously}{April 23, 2023}{Hamburg, Germany}
\acmBooktitle{CHI '23 Workshop on Designing Technology and Policy Simultaneously, April 23, 2023, Hamburg, Germany}

\usepackage[suppress]{color-edits}
\usepackage{comment}
\usepackage{color-edits}
\addauthor{ak}{brown} 
\addauthor{hz}{pink} 
\addauthor{kh}{purple} 
\addauthor{hh}{orange} 
\addauthor{ac}{red} 
\addauthor{placeholder}{red}

\begin{document}

\title{Recentering Validity Considerations through\\Early-Stage Deliberations Around AI and Policy Design}

\author{Anna Kawakami}
\affiliation{%
  \institution{Carnegie Mellon University}
  \city{Pittsburgh}
  \country{USA}
}
\email{akawakam@andrew.cmu.edu}

\author{Amanda Coston}
\affiliation{%
  \institution{Carnegie Mellon University}
  \city{Pittsburgh}
  \country{USA}
}
\email{acoston@cs.cmu.edu}

\author{Haiyi Zhu}
\authornote{Co-senior authors contributed equally to this research.}
\affiliation{%
  \institution{Carnegie Mellon University}
  \city{Pittsburgh}
  \country{USA}
}
\email{haiyiz@cs.cmu.edu}

\author{Hoda Heidari}
\authornotemark[1]
\affiliation{%
  \institution{Carnegie Mellon University}
  \city{Pittsburgh}
  \country{USA}
}
\email{hheidari@cs.cmu.edu}

\author{Kenneth Holstein}
\authornotemark[1]
\affiliation{%
  \institution{Carnegie Mellon University}
  \city{Pittsburgh}
  \country{USA}
}
\email{kenneth.holstein@gmail.com}

\renewcommand{\shortauthors}{Anna Kawakami et. al.}

\begin{abstract}
AI-based decision-making tools are rapidly spreading across a range of real-world, complex domains like healthcare, criminal justice, and child welfare. A growing body of research has called for increased scrutiny around the \textit{validity} of AI system designs. However, in real-world settings, it is often not possible to fully address questions around the validity of an AI tool without also considering the design of associated organizational and public policies. Yet, considerations around how an AI tool may interface with policy are often only discussed retrospectively, \textit{after} the tool is designed or deployed. In this short position paper, we discuss opportunities to promote multi-stakeholder deliberations around the design of AI-based technologies and associated policies, at the earliest stages of a new project.
\end{abstract}




\maketitle

\section{Motivation}
Organizations are rapidly adopting AI-based decision-making tools to augment human expert decisions in high-stakes settings like child maltreatment, criminal justice, and healthcare~\cite{De-Arteaga2021, Holstein, yang2016investigating}. Research and development efforts around these tools have aimed to help overcome resource constraints and limitations \hhedit{(such as \khedit{inconsistencies and cognitive biases}\khdelete{inconsistency and certain psychological biases)}} in human decision-making~\cite{Chouldechova2018, kahneman2021noise, Levy2021}. However, the in-\hhedit{situ} use of AI-based decision-making tools has been met with significant contention~\cite{De-Arteaga2020, Green2019, Holstein, Levy2021, HoltenMoller2020}. A growing body of research and media has surfaced ways in which AI-based decision-making tools have failed to produce value in practice, despite showing promising evaluation results prior to deployment \cite{Yang2019, kawakami2022improving}. To address these concerns, research and policymaking efforts have increasingly focused on improving \hhedit{the downstream} properties \hhedit{of AI models} such as fairness, interpretability, or predictive accuracy\akdelete{ in deployments}. These efforts often begin with the assumption that the AI tool actually ``works'' and that its design is basically sound, apart from such concerns~\cite{coston2022validity, raji2022fallacy}. 
\akdelete{This growing body of research and policy effort often starts from the assumption that the AI tool actually “works” and that its design is basically sound, apart from such fairness or interpertability-related concerns~\cite{coston2022validity, raji2022fallacy}. }

However, field studies of actual AI-based decision-making tools used in organizations today are beginning to surface fundamental challenges around the \textit{validity} of \khedit{these tools}\khdelete{the underlying model} (e.g., whether the model does what it purports to do). In \khedit{complex, real-world}\khdelete{many real-world complex} decision-making contexts\akdelete{ (e.g., child welfare, criminal justice, education)}, models are typically trained to predict an imperfect proxy for \khdelete{the }human decision-maker\khedit{s'}\khdelete{’s} actual decision-making goal\khedit{s}. For example, in child welfare, prior research discussed how frontline workers are required to make day-to-day decisions with an AI tool that predicted outcomes \khedit{misaligned}\khdelete{incompatible} with their actual decision-making objectives, professional training, and legal constraints~\cite{kawakami2022improving}. 
While child welfare workers consider the\textit{ immediate safety risks and harms} to a child to make decisions about screening investigation, the dominant design for an AI tool in this domain uses \textit{long-term predictions of child placement out of their home}. In healthcare, clinicians may make decisions about resource allocation for high-risk patients by assessing each patient’s\textit{ immediate medical needs}, while an AI tool may predict \textit{longer-term healthcare costs}~\cite{kerr1975folly}. In these decision-making contexts, underlying model validity and value alignment challenges have far-reaching downstream impacts on broader organizational culture and community welfare~\cite{brown2019toward, cheng2021soliciting}. \akdelete{While many of these decision-making contexts operate within}\akedit{While expert decisions in these settings are guided by considerations around} existing legal system\hhedit{s}, developers' current processes for designing models \khedit{often fail to meaningfully}\khdelete{seldom} involve \khdelete{expert decision-makers,}legal experts, \khdelete{or }policymakers\khedit{, or decision-makers with direct expertise}. 
Instead, considerations around how an AI tool may interface with policy are often discussed retrospectively, \textit{after} the tool is designed or \hhedit{deployed}~\cite{jackson2014policy,yang2023designing}. 

This status quo design process, \hhdelete{delineating}\hhedit{scattering\akdelete{(?)}} policy and design considerations across time and space, presents several challenges to ensuring the \khdelete{valid and functional }design of \akedit{sufficiently} \khedit{valid\akdelete{and functional}} AI \khedit{tools}\khdelete{models} in real-world social decision contexts. In many cases, \textbf{it is not possible to fully address questions about \khedit{the}\khdelete{model} validity \khedit{of an AI tool} without also \khedit{considering the design of}\khdelete{understanding, integrating, or creating} both organizational and public policies} that shape \khedit{how the system will be used}\khdelete{the design or use of the system}.
For example, evaluating \khedit{whether}\khdelete{how well} a \akedit{design proposal for an} AI\khedit{-based risk assessment tool} \khdelete{model's predictive target}captures an appropriate notion of\khdelete{captures appropriate notions of child maltreatment} ``risk'' requires \hhedit{understanding \khdelete{the }legal \khedit{definitions of}\khdelete{definition of} ``risk'' and relevant \akedit{policies governing how frontline decision-makers currently make decisions}\akdelete{ and prior rulings}}. 
\akedit{Without considering \khedit{interactions between technology design, law, and policy in the \khedit{process of designing an AI tool's objective function}\khdelete{design process}, the resulting AI tool may}\khdelete{the legal system or policies that shape frontline workers' decision-making tasks or goals, frontline workers may observe that the AI model's notions of ``risk''} lack \akdelete{face}validity.}
\akedit{In this case, early-stage conversations around policy and design could proactively prompt new evaluations that assess the face or construct validity of \khedit{a proposed AI tool design}, in the context of proposed or existing organizational policies.\khdelete{an AI model}\akdelete{. These insights could also inform the design of organizational policies empowering frontline workers to provide post-deployment feedback on changes to model validity over time. }} 
Beyond this specific example, there is a broader missed opportunity for communities of stakeholders (e.g., policymakers, frontline workers, leadership, developers) to proactively exchange and synthesize knowledge around validity to make design decisions that are both informed by, and inform, policy. 






\section{Centering Validity in Early-Stage AI and Policy Design Deliberations} 
Properly addressing these challenges requires turning our attention to the earliest stages of model development and adoption: \textbf{How can we refocus policy and research efforts around \akdelete{functionality and }validity considerations, by promoting early-stage deliberations around \khedit{how to design AI-based technologies and associated policies?}\khdelete{whether and how to design improved technologies and policies?} }Today, we lack effective, practical processes for proactively engaging policymakers, developers, and other stakeholders in fundamental questions around the design and governance of AI systems (e.g., whether a deployed AI system will actually do what it purports to do). In this position paper, we propose supporting early-stage, multi-stakeholder deliberations around the \akedit{validity}\akdelete{functionality} of proposed AI tools as a step towards designing better policies and technologies together. 

A growing chorus of research has called for increased developer attention around AI validity concerns \akedit{(sometimes discussed \khedit{via related concepts such} as ``AI functionality'')} as an essential first step towards ensuring the safety of AI deployments~\cite{coston2022validity, raji2022fallacy, wang2022against}. However, much of this work still lives at the theoretical level, geared towards academic researchers. Grounding these considerations around \akdelete{functionality and }validity into real-world design and policymaking settings requires a diverse pool of expertise – from an understanding of existing organizational processes, needs, and constraints around designing AI tools to tacit knowledge of opportunities and boundaries for informing policy. Relatedly, it is critical that such early-stage deliberations promote knowledge-sharing and synthesis across a wide range of relevant stakeholders, including policymakers, frontline workers, community members, developers, and organizational leaders. Through this \hhedit{piece, we invite \akedit{researchers, designers, and practitioners}\akdelete{the Human-AI interaction community}} to explore anticipated challenges and opportunities to operationalizing these multi-stakeholder early-stage deliberations for policy and design.


\section{Open Questions}
We invite the human-AI interaction, science and technology studies, machine learning, and other relevant communities to further knowledge around these topics: 
\akdelete{While this direction is still nascent, we briefly begin a discussion on open questions we hope to continue exploring through the workshop:}

\textbf{Shifting power imbalances in and through collaborative design.}
Effective early-stage deliberations around policy and design requires engaging stakeholders (e.g., frontline decision-makers and community members) who are often left out of the model design process in the current status quo. In other words, implementing an effective deliberation process may also require shifting institutional power imbalances across stakeholders of the AI tool. At the same time, a deliberation process in itself may be structured to intentionally help shift power imbalances, for example, through the use of accessible and shared language or stakeholder-specific questions. How can we best shift power imbalances across different stakeholders varying in position and background, in the process of collaboratively designing policy and technology? What forms of imbalances cannot be nudged through deliberations, and how might other forms of support play a role?  

\textbf{Supporting evaluative and generative discourse.} Deliberations about the validity of AI tools may need to be both evaluative and generative, to promote sufficient organizational buy-in and ensure resulting ideas produce more benefits than harms in practice. However, there may be tensions between different stakeholders and their (perceived) stances towards technical innovation versus evaluation. For example, there may be a (mis)conception that policies and laws constrain technical innovation, hindering effective conversation. How can we promote shared goals for collaborative policy and design discussions, while also valuing and leveraging differences in perspectives towards the role of AI innovation in a given context? 

\textbf{Connecting to local policymaking organizations.}
While early-stage deliberations may help identify and design new policies to complement technology design, they do not ensure that such policies are actually implemented post-deliberation. How can we better connect organizations with local policymaking groups, to help streamline outputs resulting from designing policy and technology together? 

\textbf{Overcoming incentive structures and pressures.}
In practice, there may be incentive structures, social pressures, or infrastructural barriers that hinder different stakeholders’ desires or abilities to engage in substantive discourse around designing more responsible technologies and policies. What role might other forces of higher power (e.g., regulation) play in ensuring that process-oriented solutions like supporting early-stage deliberations have sufficient teeth in practice? 

\textbf{Exploring other design opportunities for recentering validity.}
Supporting early-stage deliberations is one possible solution for recentering validity considerations in policy and design discourse. However, it may not be the best solution. What might other opportunities for refocusing validity considerations in design and policy look like?

\begin{acks}
This work was generously funded by CMU Block Center for Technology and Society Award No. 53680.1.5007718.
\end{acks}

\bibliographystyle{ACM-Reference-Format}
\bibliography{functpolicydesign}


\begin{thebibliography}{19}


\ifx \showCODEN    \undefined \def \showCODEN     #1{\unskip}     \fi
\ifx \showDOI      \undefined \def \showDOI       #1{#1}\fi
\ifx \showISBNx    \undefined \def \showISBNx     #1{\unskip}     \fi
\ifx \showISBNxiii \undefined \def \showISBNxiii  #1{\unskip}     \fi
\ifx \showISSN     \undefined \def \showISSN      #1{\unskip}     \fi
\ifx \showLCCN     \undefined \def \showLCCN      #1{\unskip}     \fi
\ifx \shownote     \undefined \def \shownote      #1{#1}          \fi
\ifx \showarticletitle \undefined \def \showarticletitle #1{#1}   \fi
\ifx \showURL      \undefined \def \showURL       {\relax}        \fi
\providecommand\bibfield[2]{#2}
\providecommand\bibinfo[2]{#2}
\providecommand\natexlab[1]{#1}
\providecommand\showeprint[2][]{arXiv:#2}

\bibitem[Brown et~al\mbox{.}(2019)]%
        {brown2019toward}
\bibfield{author}{\bibinfo{person}{Anna Brown}, \bibinfo{person}{Alexandra
  Chouldechova}, \bibinfo{person}{Emily Putnam-Hornstein},
  \bibinfo{person}{Andrew Tobin}, {and} \bibinfo{person}{Rhema Vaithianathan}.}
  \bibinfo{year}{2019}\natexlab{}.
\newblock \showarticletitle{Toward algorithmic accountability in public
  services: A qualitative study of affected community perspectives on
  algorithmic decision-making in child welfare services}. In
  \bibinfo{booktitle}{\emph{Proceedings of the 2019 CHI Conference on Human
  Factors in Computing Systems}}. \bibinfo{pages}{1--12}.
\newblock


\bibitem[Cheng et~al\mbox{.}(2021)]%
        {cheng2021soliciting}
\bibfield{author}{\bibinfo{person}{Hao-Fei Cheng}, \bibinfo{person}{Logan
  Stapleton}, \bibinfo{person}{Ruiqi Wang}, \bibinfo{person}{Paige Bullock},
  \bibinfo{person}{Alexandra Chouldechova}, \bibinfo{person}{Zhiwei
  Steven~Steven Wu}, {and} \bibinfo{person}{Haiyi Zhu}.}
  \bibinfo{year}{2021}\natexlab{}.
\newblock \showarticletitle{Soliciting stakeholders’ fairness notions in
  child maltreatment predictive systems}. In
  \bibinfo{booktitle}{\emph{Proceedings of the 2021 CHI Conference on Human
  Factors in Computing Systems}}. \bibinfo{pages}{1--17}.
\newblock


\bibitem[Chouldechova et~al\mbox{.}(2018)]%
        {Chouldechova2018}
\bibfield{author}{\bibinfo{person}{Alexandra Chouldechova},
  \bibinfo{person}{Emily Putnam-Hornstein}, \bibinfo{person}{Suzanne
  Dworak-Peck}, \bibinfo{person}{Diana Benavides-Prado},
  \bibinfo{person}{Oleksandr Fialko}, \bibinfo{person}{Rhema Vaithianathan},
  \bibinfo{person}{Sorelle~A Friedler}, {and} \bibinfo{person}{Christo
  Wilson}.} \bibinfo{year}{2018}\natexlab{}.
\newblock \showarticletitle{{A case study of algorithm-assisted decision making
  in child maltreatment hotline screening decisions}}.
\newblock \bibinfo{journal}{\emph{Proceedings of Machine Learning Research}}
  \bibinfo{volume}{81} (\bibinfo{year}{2018}), \bibinfo{pages}{1--15}.
\newblock
\showISSN{2640-3498}
\urldef\tempurl%
\url{http://proceedings.mlr.press/v81/chouldechova18a.html}
\showURL{%
\tempurl}


\bibitem[Coston et~al\mbox{.}(2022)]%
        {coston2022validity}
\bibfield{author}{\bibinfo{person}{Amanda Coston}, \bibinfo{person}{Anna
  Kawakami}, \bibinfo{person}{Haiyi Zhu}, \bibinfo{person}{Ken Holstein}, {and}
  \bibinfo{person}{Hoda Heidari}.} \bibinfo{year}{2022}\natexlab{}.
\newblock \showarticletitle{A Validity Perspective on Evaluating the Justified
  Use of Data-driven Decision-making Algorithms}.
\newblock \bibinfo{journal}{\emph{arXiv preprint arXiv:2206.14983}}
  (\bibinfo{year}{2022}).
\newblock


\bibitem[De-Arteaga et~al\mbox{.}(2021)]%
        {De-Arteaga2021}
\bibfield{author}{\bibinfo{person}{Maria De-Arteaga}, \bibinfo{person}{Artur
  Dubrawski}, {and} \bibinfo{person}{Alexandra Chouldechova}.}
  \bibinfo{year}{2021}\natexlab{}.
\newblock \showarticletitle{{Leveraging expert consistency to improve
  algorithmic decision support}}.
\newblock \bibinfo{journal}{\emph{arXiv}} (\bibinfo{year}{2021}),
  \bibinfo{pages}{1--33}.
\newblock
\showeprint[arxiv]{2101.09648}
\urldef\tempurl%
\url{http://arxiv.org/abs/2101.09648}
\showURL{%
\tempurl}


\bibitem[De-Arteaga et~al\mbox{.}(2020)]%
        {De-Arteaga2020}
\bibfield{author}{\bibinfo{person}{Maria De-Arteaga}, \bibinfo{person}{Riccardo
  Fogliato}, {and} \bibinfo{person}{Alexandra Chouldechova}.}
  \bibinfo{year}{2020}\natexlab{}.
\newblock \showarticletitle{{A case for humans-in-the-loop: Decisions in the
  presence of erroneous algorithmic scores}}.
\newblock \bibinfo{journal}{\emph{arXiv}} (\bibinfo{year}{2020}),
  \bibinfo{pages}{1--12}.
\newblock
\showISBNx{9781450367080}
\showISSN{23318422}


\bibitem[Green and Chen(2019)]%
        {Green2019}
\bibfield{author}{\bibinfo{person}{Ben Green} {and} \bibinfo{person}{Yiling
  Chen}.} \bibinfo{year}{2019}\natexlab{}.
\newblock \showarticletitle{{The principles and limits of algorithm-in-the-loop
  decision making}}.
\newblock \bibinfo{journal}{\emph{Proceedings of the ACM on Human-Computer
  Interaction}} \bibinfo{volume}{3}, \bibinfo{number}{CSCW}
  (\bibinfo{year}{2019}).
\newblock
\showISSN{25730142}
\urldef\tempurl%
\url{https://doi.org/10.1145/3359152}
\showDOI{\tempurl}


\bibitem[Holstein and Aleven(2021)]%
        {Holstein}
\bibfield{author}{\bibinfo{person}{Kenneth Holstein} {and}
  \bibinfo{person}{Vincent Aleven}.} \bibinfo{year}{2021}\natexlab{}.
\newblock \showarticletitle{Designing for human-AI complementarity in K-12
  education}.
\newblock \bibinfo{journal}{\emph{arXiv preprint arXiv:2104.01266}}
  (\bibinfo{year}{2021}).
\newblock


\bibitem[{Holten M{\o}ller} et~al\mbox{.}(2020)]%
        {HoltenMoller2020}
\bibfield{author}{\bibinfo{person}{Naja {Holten M{\o}ller}},
  \bibinfo{person}{Irina Shklovski}, {and} \bibinfo{person}{Thomas~T.
  Hildebrandt}.} \bibinfo{year}{2020}\natexlab{}.
\newblock \showarticletitle{{Shifting concepts of value: Designing algorithmic
  decision-support systems for public services}}.
\newblock \bibinfo{journal}{\emph{NordiCHI}} (\bibinfo{year}{2020}),
  \bibinfo{pages}{1--12}.
\newblock
\showISBNx{9781450375795}
\urldef\tempurl%
\url{https://doi.org/10.1145/3419249.3420149}
\showDOI{\tempurl}


\bibitem[Jackson et~al\mbox{.}(2014)]%
        {jackson2014policy}
\bibfield{author}{\bibinfo{person}{Steven~J Jackson}, \bibinfo{person}{Tarleton
  Gillespie}, {and} \bibinfo{person}{Sandy Payette}.}
  \bibinfo{year}{2014}\natexlab{}.
\newblock \showarticletitle{The policy knot: Re-integrating policy, practice
  and design in CSCW studies of social computing}. In
  \bibinfo{booktitle}{\emph{Proceedings of the 17th ACM conference on Computer
  supported cooperative work \& social computing}}. \bibinfo{pages}{588--602}.
\newblock


\bibitem[Kahneman et~al\mbox{.}(2021)]%
        {kahneman2021noise}
\bibfield{author}{\bibinfo{person}{Daniel Kahneman}, \bibinfo{person}{Olivier
  Sibony}, {and} \bibinfo{person}{Cass~R Sunstein}.}
  \bibinfo{year}{2021}\natexlab{}.
\newblock \bibinfo{booktitle}{\emph{Noise: A flaw in human judgment}}.
\newblock \bibinfo{publisher}{Little, Brown}.
\newblock


\bibitem[Kawakami et~al\mbox{.}(2022)]%
        {kawakami2022improving}
\bibfield{author}{\bibinfo{person}{Anna Kawakami}, \bibinfo{person}{Venkatesh
  Sivaraman}, \bibinfo{person}{Hao-Fei Cheng}, \bibinfo{person}{Logan
  Stapleton}, \bibinfo{person}{Yanghuidi Cheng}, \bibinfo{person}{Diana Qing},
  \bibinfo{person}{Adam Perer}, \bibinfo{person}{Zhiwei~Steven Wu},
  \bibinfo{person}{Haiyi Zhu}, {and} \bibinfo{person}{Kenneth Holstein}.}
  \bibinfo{year}{2022}\natexlab{}.
\newblock \showarticletitle{Improving Human-AI Partnerships in Child Welfare:
  Understanding Worker Practices, Challenges, and Desires for Algorithmic
  Decision Support}. In \bibinfo{booktitle}{\emph{CHI Conference on Human
  Factors in Computing Systems}}. \bibinfo{pages}{1--18}.
\newblock


\bibitem[Kerr(1975)]%
        {kerr1975folly}
\bibfield{author}{\bibinfo{person}{Steven Kerr}.}
  \bibinfo{year}{1975}\natexlab{}.
\newblock \showarticletitle{On the folly of rewarding A, while hoping for B}.
\newblock \bibinfo{journal}{\emph{Academy of Management journal}}
  \bibinfo{volume}{18}, \bibinfo{number}{4} (\bibinfo{year}{1975}),
  \bibinfo{pages}{769--783}.
\newblock


\bibitem[Levy et~al\mbox{.}(2021)]%
        {Levy2021}
\bibfield{author}{\bibinfo{person}{Karen Levy}, \bibinfo{person}{Kyla~E
  Chasalow}, {and} \bibinfo{person}{Sarah Riley}.}
  \bibinfo{year}{2021}\natexlab{}.
\newblock \showarticletitle{{Algorithms and Decision-Making in the Public
  Sector}}.
\newblock \bibinfo{journal}{\emph{Annual Review of Law and Social Science}}
  \bibinfo{volume}{17} (\bibinfo{year}{2021}), \bibinfo{pages}{1--38}.
\newblock


\bibitem[Raji et~al\mbox{.}(2022)]%
        {raji2022fallacy}
\bibfield{author}{\bibinfo{person}{Inioluwa~Deborah Raji},
  \bibinfo{person}{I~Elizabeth Kumar}, \bibinfo{person}{Aaron Horowitz}, {and}
  \bibinfo{person}{Andrew Selbst}.} \bibinfo{year}{2022}\natexlab{}.
\newblock \showarticletitle{The fallacy of AI functionality}. In
  \bibinfo{booktitle}{\emph{2022 ACM Conference on Fairness, Accountability,
  and Transparency}}. \bibinfo{pages}{959--972}.
\newblock


\bibitem[Wang et~al\mbox{.}(2022)]%
        {wang2022against}
\bibfield{author}{\bibinfo{person}{Angelina Wang}, \bibinfo{person}{Sayash
  Kapoor}, \bibinfo{person}{Solon Barocas}, {and} \bibinfo{person}{Arvind
  Narayanan}.} \bibinfo{year}{2022}\natexlab{}.
\newblock \showarticletitle{Against Predictive Optimization: On the Legitimacy
  of Decision-Making Algorithms that Optimize Predictive Accuracy}.
\newblock \bibinfo{journal}{\emph{Available at SSRN}} (\bibinfo{year}{2022}).
\newblock


\bibitem[Yang et~al\mbox{.}(2019)]%
        {Yang2019}
\bibfield{author}{\bibinfo{person}{Qian Yang}, \bibinfo{person}{Aaron
  Steinfeld}, {and} \bibinfo{person}{John Zimmerman}.}
  \bibinfo{year}{2019}\natexlab{}.
\newblock \showarticletitle{{Unremarkable AI: Fitting intelligent decision
  support into critical, clinical decision-making processes}}.
\newblock \bibinfo{journal}{\emph{Conference on Human Factors in Computing
  Systems - Proceedings}} (\bibinfo{year}{2019}).
\newblock
\showISBNx{9781450359702}
\urldef\tempurl%
\url{https://doi.org/10.1145/3290605.3300468}
\showDOI{\tempurl}
\showeprint[arxiv]{1904.09612}


\bibitem[Yang et~al\mbox{.}(2023)]%
        {yang2023designing}
\bibfield{author}{\bibinfo{person}{Qian Yang}, \bibinfo{person}{Richmond Wong},
  \bibinfo{person}{Thomas Gilbert}, \bibinfo{person}{Margaret Hagan},
  \bibinfo{person}{Steven Jackson}, \bibinfo{person}{Sabine Junginger}, {and}
  \bibinfo{person}{John Zimmerman}.} \bibinfo{year}{2023}\natexlab{}.
\newblock \showarticletitle{Designing Technology and Policy Simultaneously:
  Towards A Research Agenda and New Practice}.
\newblock  (\bibinfo{year}{2023}).
\newblock


\bibitem[Yang et~al\mbox{.}(2016)]%
        {yang2016investigating}
\bibfield{author}{\bibinfo{person}{Qian Yang}, \bibinfo{person}{John
  Zimmerman}, \bibinfo{person}{Aaron Steinfeld}, \bibinfo{person}{Lisa Carey},
  {and} \bibinfo{person}{James~F Antaki}.} \bibinfo{year}{2016}\natexlab{}.
\newblock \showarticletitle{Investigating the heart pump implant decision
  process: opportunities for decision support tools to help}. In
  \bibinfo{booktitle}{\emph{Proceedings of the 2016 CHI Conference on Human
  Factors in Computing Systems}}. \bibinfo{pages}{4477--4488}.
\newblock


\end{thebibliography}


\end{document}